\documentclass[ aip,reprint]{revtex4-1}

\usepackage{graphics} 
\usepackage{epsfig} 
\usepackage{amsmath} 
\usepackage{amssymb} 
\usepackage{commath}
\usepackage{amsthm}

\usepackage{dcolumn}

\usepackage[T1]{fontenc}
\usepackage{mathptmx}

\usepackage{amsfonts}
\usepackage{bm} 
\usepackage{graphicx}
\usepackage[usenames,dvipsnames,table]{xcolor}

\def\bF{\bm{F}}

\def\bx{\bm{x}}

\def\br{\mathbf{r}}
\def\bR{\bm{R}}
\def\R{\mb{R}}
\def\>{\rightarrow}
\def\({\left(}
\def\){\right)}

\def\>{\rightarrow}
\def\grad{\nabla}

\newcommand{\mb}[1]{\bm{{#1}}}
\newcommand{\mc}[1]{\mathcal{{#1}}}

\newcommand{\DS}{\displaystyle}

\begin{document}

\title{Torus bifurcations of large-scale swarms having range dependent communication delay}

\author{Ira B. Schwartz}
\email{ira.schwartz@nrl.navy.mil}
\affiliation{U.S. Naval Research Laboratory, Code 6792, Plasma Physics
  Division, Washington, DC 20375, USA}
\author{Victoria Edwards}
\affiliation{U.S. Naval Research Laboratory, Code 5514, Navy Center for Applied Research in Artificial Intelligence, Washington, DC 20375, USA}
\author{Sayomi Kamimoto}
\affiliation{Department of Mathematics, George Mason University, Fairfax
  Virginia, 22030, USA}
\author{Klimka Kasraie}
\affiliation{Aerospace, Transportation and Advanced Systems Laboratory of the Georgia Tech Research Institute, Atlanta, GA 30332}

\author{M. Ani Hsieh}
\affiliation{Mechanical Engineering and Applied Mechanics
  University of Pennsylvania, Philadelphia, PA 19104 USA}
\author{Ioana Triandaf$^1$}
\author{Jason Hindes$^1$}
\date{\today}

\begin{abstract}
Dynamical emergent  patterns of swarms are now fairly well established in nature, and
include flocking  and rotational states.  Recently,
there has been  great interest in engineering and physics to create artificial
self-propelled agents that communicate over a network and operate with simple rules, with the goal
of creating emergent self-organizing swarm patterns. In this paper, we show that when
communicating networks have  range dependent delays, rotational states
which are typically periodic, undergo a bifurcation and create swarm dynamics on
a torus. The observed bifurcation yields additional frequencies into the
dynamics, which may lead to quasi-periodic behavior of the swarm. 
\end{abstract}  

\maketitle
\textbf{    Swarming behavior occurs when a large number of self-propelled agents interact using simple rules.
    Natural swarms of biological systems have been observed at a range of length scales forming complex emergent patterns.
    Engineers have drawn inspiration from these natural systems, resulting in the translation of swarm theory to communicating robotic systems.
    Example applications of artificial swarms include: exploration and mapping, search and rescue, and distributed sensing and estimation.
    Through continued development, an additional parameter of delay in communication between artificial agents has become important to consider. 
    Specifically, it was previously discovered, that communication delay will create new rotational patterns which are not observed without delay, both theoretically and experimentally.
    Here we extend the understanding of communication delays to reveal the effects of range dependent delay, where the communication between agents depends
    on the distance between agents.
    The results of the research show that by including range dependent delay, new rotational states are introduced.
    We show how these new states emerge, discuss their stability, and discuss
    how they may be realized in large scale robotic systems. In improving our theoretical
  understanding of predicted swarm behavior modeled in simulation we can
  better anticipate what will happen experimentally. Additionally, it is
  possible to leverage the predicted autonomous behaviors to try and force different swarm behavior.}

\section{Introduction}
Swarming behavior, which we define as the emergence of spatio-temporal group behaviors from simple local interactions between pairs of agents, is  widespread and observed over a range of application domains. Examples
can be found in biological systems over a range of length scales, from   aggregates of bacterial cells and
dynamics of skin cells in wound healing \cite{Budrene1995,Polezhaev2006,Lee2013} to
dynamic patterns of fish, birds, bats, and even humans
\cite{Tunstrom2013,Helbing1995,Giuggioli2015,Lee2006}. 
These systems are particularly interesting  because they allow simple
individual agents to achieve complex tasks in ways that are scalable, extensible, and robust to
failures of individual agents. In addition, these swarming behaviors are able to form and persist
in spite of complicating factors such as  delayed actuation, latent
communication, localized  number of
neighbors each agent is able to interact with, heterogeneity in agent dynamics, and environmental
noise. These factors have been the focus of previous theoretical research in describing
the bifurcating spatial-temporal patterns in swarms, as seen for example in
Refs. 
\cite{Topaz2004,Szwaykowska2014,Romero2011,Hindes2016}. Likewise, the application of swarms have been experimentally realized in areas,  such as
mapping\cite{Ramachandran2018}, leader-following\cite{MORGAN2005,Wiech2018},
and density control\cite{Li17}. To guarantee swarming behavior experimentally,  
control is typically employed \cite{Tanner07, Gazi05, Jadbabaie03, Viragh2014,
  Desai01} to prove convergence to a given state by  relying on
strict assumptions to guarantee the desired behavior. However, by relaxing certain assumptions, a  number of studies show that even with simple interaction protocols, swarms of agents are able to
converge to organized, coherent behaviors in a self-emergent manner;
i.e. autonomously  without
control. Different mathematical approaches
{yielded} a  wide
selection of both agent-based
\cite{Helbing1995,Lee2006,Vicsek2006,Tunstrom2013} and continuum models
that predict  swarming dynamics.
\cite{Edelstein-Keshet1998,Topaz2004,Polezhaev2006}. In almost all models,
since the agents have just a few simple rules, there exists only a relatively
small number of controllable parameters. The parameter set usually consists of a
self-propulsion force, a potential function governing attracting and repelling
forces between agents, and a communicating radius governing the local
neighborhood at which the agents can sense and  interact with each other.

In both robotic and biological swarms, an additional parameter appears as a
delay between the time information is perceived and the actuation (reaction) time of an agent. 
Such delays have now  been measured in swarms of bats, 
birds, fish, and crowds of
people\cite{Luca_bats,Nagy_pigeons,JF_people}. The measured delays are
longer than the typical relaxation times of the agents, and may be space and
time dependent. Robotic swarms experience communication delays which provide similar effects to the delay experienced in natural swarms. Incorporating stationary delays
along with a minimal set of parameters in swarm models results in multi-stability of
rotational patterns in space~\cite{Romero2012, Szwaykowska2016, Edwards2019, 
   Hindes18, Szwaykowska2018}. 
In particular, for delays that equal and fixed, one observes three basic
 swarming states or modes: Flocking, which
  is a translating center of mass, Ring state, where the agents are splayed
  out on a ring in in phase about a stationary center of mass, and a Rotating
  state, where the center of mass itself rotates.

Synthetic robotic swarms have communication delays that naturally
occur over wireless networks, as a  result of low
bandwidth\cite{Komareji2018} resulting in delayed communication and multi-hop communication\cite{Oliveira15}. In
cases where the delays are fixed and equal, and the communication occurs on a
homogeneous network, it is known that delays create new rotational patterns
, as has been verified both theoretically and experimentally
\cite{Szwaykowska2016, Edwards2019}. However, in situations with robots, even simple communication models are based on the distance between agents \cite{Hsieh04, Hsieh07}.
Following from these models, if one assumes that the delays are range dependent, the problem becomes one of studying state dependent delays where delays depend implicitly on the relative
positions between agents.

{When placing swarms in realistic complex environments,  delays are not necessarily
a continuous function of range, but rather it is the increasing probability of
delays increasing stochastically when agents move further away from one
another beyond a certain radius~\cite{Fink2012,Fink2013}. That is, the rate of communication becomes spatially
dependent, whereby near agents see a signal with a fast rate of communication,
but due to shading and fading of signals, communication rates are slowed and complex
outside a given radius. Underwater communication is an excellent swarm example
where delays outside a significant radius impart rates of communication of one
to two 
orders of magnitude greater than local communication rates~\cite{UnderwaterComms2009}.}

The swarm model  that follows takes a globally
coupled swarm, and explicitly relaxes the fixed delay assumption, by including
range dependent delay based on a fixed communication radius. We show that when range dependent delays are included,
new frequencies are introduced and generate bifurcations to a torus. The result
is a milling type of swarm that depends on just a few parameters.  The results here are important for robotic swarming where one of the goals is to produce desired patterns autonomously, without external controls. The pattern formations predicted here show how delayed information, whether coming from communication, actuations, or both, impacts the stability of swarm states, such as ring and/or rotating states. By revealing those parameter regions where patterns are destabilized, we provide a comprehensive characterization of the autonomously accessible swarm states in the presence of range-dependent delay. 

\section{The swarm model}
Consider a swarm of delay-coupled agents in $\R^2$.  Each agent is indexed by
$i \in \{1,\ldots, N\}$. We use a simple but general model for swarming motion. Each agent has a
self-propulsion force that strives to maintain motion at a preferred speed and a coupling force that
governs its interaction with other agents in the swarm. The interaction force is defined as the
negative gradient of a pairwise interaction potential $U(\cdot,\cdot)$. All agents follow the same
rules of motion; however, mechanical differences between agents may lead to heterogeneous dynamics;
this effect is captured by assigning different acceleration factors (denoted $\kappa_i$) to the
agents. In this paper, we  assume $\kappa_i = 1$ for all $i$. For the effect
of heterogeneity on the swarm bifurcations, see \cite{Szwaykowska2014}.

Agent-to-agent interactions occur along a graph $\mc{G} = \{\mc{V},\mc{E}\}$, where $\mc{V}$ is
the set of vertexes $v_i$ in the graph and $\mc{E}$ is the set of edges $e_{ij}$. The vertices
correspond to individual swarm agents, and edges represent communication links; that is, agents $i$
and $j$ communicate with each other if and only if $e_{ij} \in \mc{E}$. All communications links are
assumed to be bi-directional, and all communications occur with a time delay
$\tau$. That is, range dependence is not included. Let ${\mb r}_i \in \mathbb{R}^2$ denote the position of the agent $i$ and let
$\mc{N}_i = \{v_j \in \mc{V} : e_{ij} \in \mc{E}\}$ denote its set of neighbors of
agent $i$.  The motion of agent $i$ is governed by the following equation:
\begin{equation} \label{Eq:agenti}
 \ddot{\br}_i = \kappa_i(1-\norm{\dot{\br}_i}^2)\dot{\br}_i - \kappa_i \sum_{j \in \mc{N}_i}\grad_x U({\br}_i(t),{\br}_j^\tau(t)),
\end{equation}
where superscript $\tau$ is used to denote time delay, so that
${\br}_j^\tau(t) = {\br}_j(t-\tau)$, $\norm{\cdot}$ denotes the Euclidean norm, and $\grad_x$
denotes the gradient with respect to the first argument of $U$. The first term in
Eq.~\ref{Eq:agenti} governs self-propulsion, where the speed has been
  normalized to unity. That is, without coupling the agents always asymptote to
unit speed.

To analyze the dynamics of a large scale swarm, we use a
harmonic interaction potential with short-range repulsion. 
\begin{equation} \label{eq:U}
 U(\br_i,\br_j^\tau) = c_r e^{-\frac{\norm{\br_i - \br_j}}{l_r}} + \frac{a}{2N}\norm{\br_i - \br_j^\tau}^2.
\end{equation}\\ 

In Eq.~\ref{Eq:agenti}, it is  assumed that the communication delay, $\tau$, is
independent of the distance, or range, between any pair of agents. (Notice that
the exponent of the repulsion term is independent of the delay since the
repulsion force is local.)  With the
addition of delays in the network, it was shown in  homogeneous communication
networks that in addition to the usual dynamical  translating and milling (or ring) states, for
sufficiently large $\tau$, new rotational states emerge \cite{Szwaykowska2016}. In particular, for a a given attractive coupling strength, there is a
delay that destabilizes the periodic  ring state into a rotating state, in which  the agents
coalesce to a small group and move around a fixed center of rotation; this
behavior is quite different from 
the ring state where agents are spread out in a splay state phase.  The rotating state is only
observed with delay introduced in the communication network, and it appears
through a Hopf bifurcation.

However, in real-world robotic swarms, communication delays are not uniform between all pairs of agents; delays may be stochastic or even state-dependent.
  For example, if agents are communicating over a multi-hop network, the delay will increase with the number of hops required to send a message from one agent to the other,
  and in general will scale with the separation between them. In order to handle range dependent
delays, we will make an approximation that depends on a communication range radius.

\subsection{Approximating range dependent delayed coupling}
For the coupling term, we are interested in introducing an approximation to
range based coupling delay. Since all communicating agents send signals
with some delay, we compute relative distances defined as 
\begin{align}
D^\tau_{i,j} &\equiv ||{\br}_i-{\br}_j^{\tau}||.
\end{align}
We define a Heaviside function, $H(x)$, that is zero when $x \le 0$
and 1 otherwise, and we employ  global coupling based on  a spring
potential. For our range dependent metric, we let $\epsilon \ge 0$ denote the range
radius. Suppose that when the separation between two agents is small, that is less than $\epsilon$,
 then sensing between two agents is almost immediate. In practice, the time
 needed for sensing depends on several
factors, such as actuation times, and so distances in practice are computed
with delay. Therefore, we model the coupling term for the $i^{th}$ agent as 


\begin{align}\label{Eq:Ci}
C_{i}({\br}_{i},{\br}_{j},{\br}_{j}^{\tau},\epsilon) = -\frac{a}{N}(\grad_x U({\br}_i(t),{\br}_j^\tau(t)))H(D^\tau_{i,j}-\epsilon) \nonumber\\
                 -\frac{a}{N}(\grad_x U({\br}_i(t),{\br}_j(t)))(1-H(D^\tau_{i,j}-\epsilon)),
\end{align}
{where the first coupling term has delay turned on since the distance is outside a ball of radius
  $\epsilon$, while the second term has no delay since the distance is within the $\epsilon$
  ball. }
The resulting swarm model with range dependence from Eq.~\ref{Eq:Ci} is now
\begin{equation} \label{Eq:agenti_range}
 \ddot{\br}_i = \kappa_i(1-\norm{\dot{\br}_i}^2)\dot{\br}_i - \kappa_i \sum_{j \in \mc{N}_i}C_{i}({\br}_{i},{\br}_{j},{\br}_{j}^{\tau},\epsilon).
\end{equation}

If the delayed distance is within an $\epsilon$ ball, then we evaluate the
coupling without delay. Otherwise the coupling is delayed. Thus the coupling
function takes into account when delay is active or not between pairs of
communicating agents, and depends on the range radius, $\epsilon$.

The Heaviside function of the  right hand side of Eq.~\ref{MF_eqn}
renders the differential delay equation derivatives discontinuous, and as such poses a numerical
integration problem. To mollify the lack of smoothness, we approximate $H(x)$
by letting $H(x) \approx \frac{1}{\pi}\arctan(kx)+\frac{1}{2}$, where $k \gg
1$ and constant, and limits on the Heaviside function as $k \rightarrow \infty$. 

Using only the delayed distance to compute a range dependent coupling assumes
that any measurement is not instantaneous. If one were to be able to compute
the ideal situation where delay would not be a sensing factor, then certain
issues would need to be resolved, which we do not consider here.\\

\subsection{ Numerical simulations of full swarms }
Examples of simulations using the swarm model with the range dependent coupling
are shown below. Here the number of agents, $N=150$, and
the coupling strength, $a=2.0$. For the remainder of the analysis, we set
$c_r=0$, and note that the attractors persist when the repulsive amplitude is
sufficiently small \cite{Szwaykowska2016}. (See supplementary material for
a video of the dynamics with small repulsion.)

\begin{center}
  \begin{figure}[h]
    \includegraphics[width=0.35\textwidth]{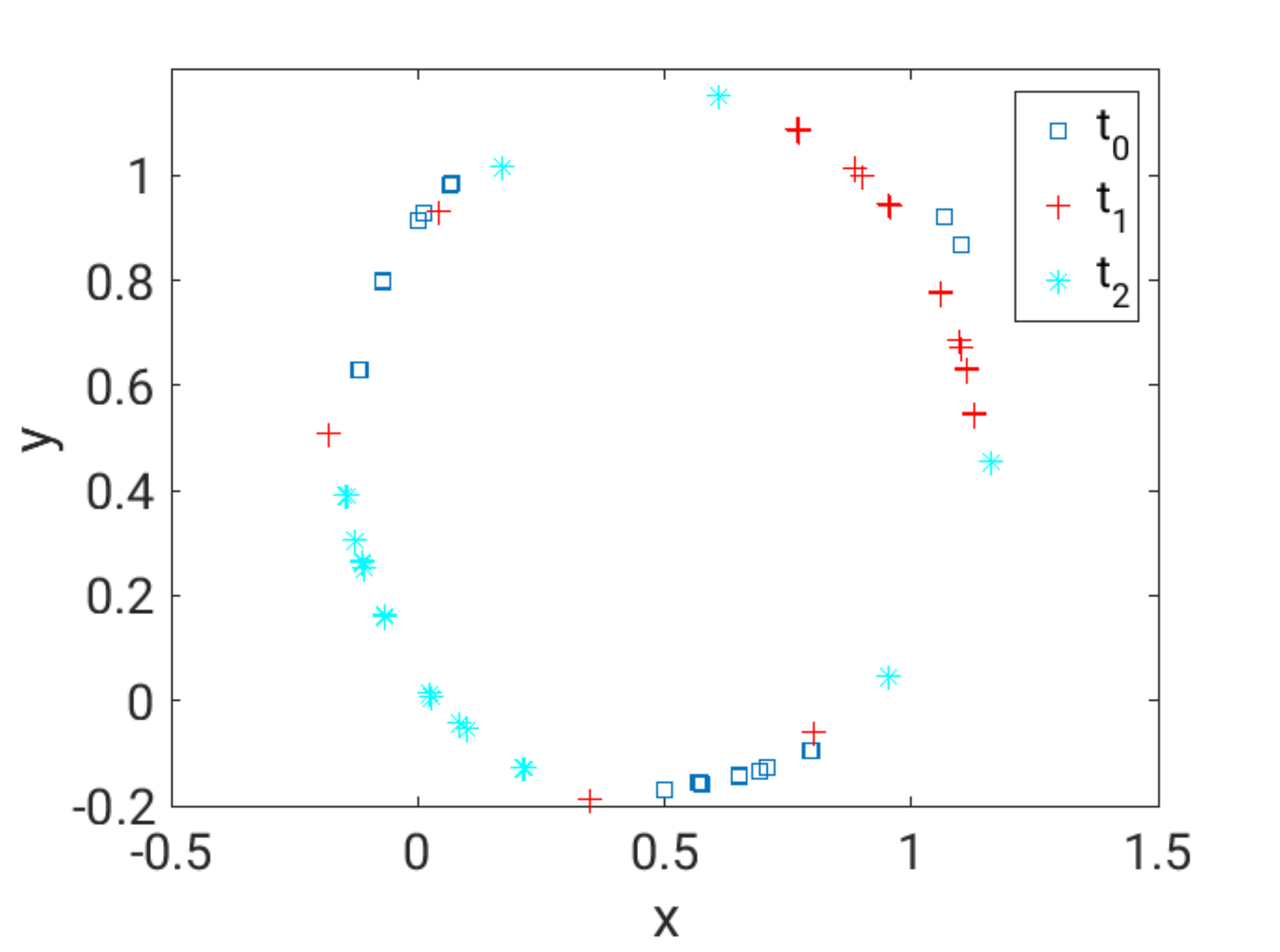}
    \caption{Three snapshots of swarm state in space for $\epsilon = 0.01,
      a=2.0, \tau = 1.75$. Sample times $t_{0},t_1=t_{0}+20,t_2=t_0+40$ .}
    \label{ThreeSnaps}
  \end{figure}
\end{center}

\begin{center}
  \begin{figure}[h]
    \includegraphics[width=0.35\textwidth]{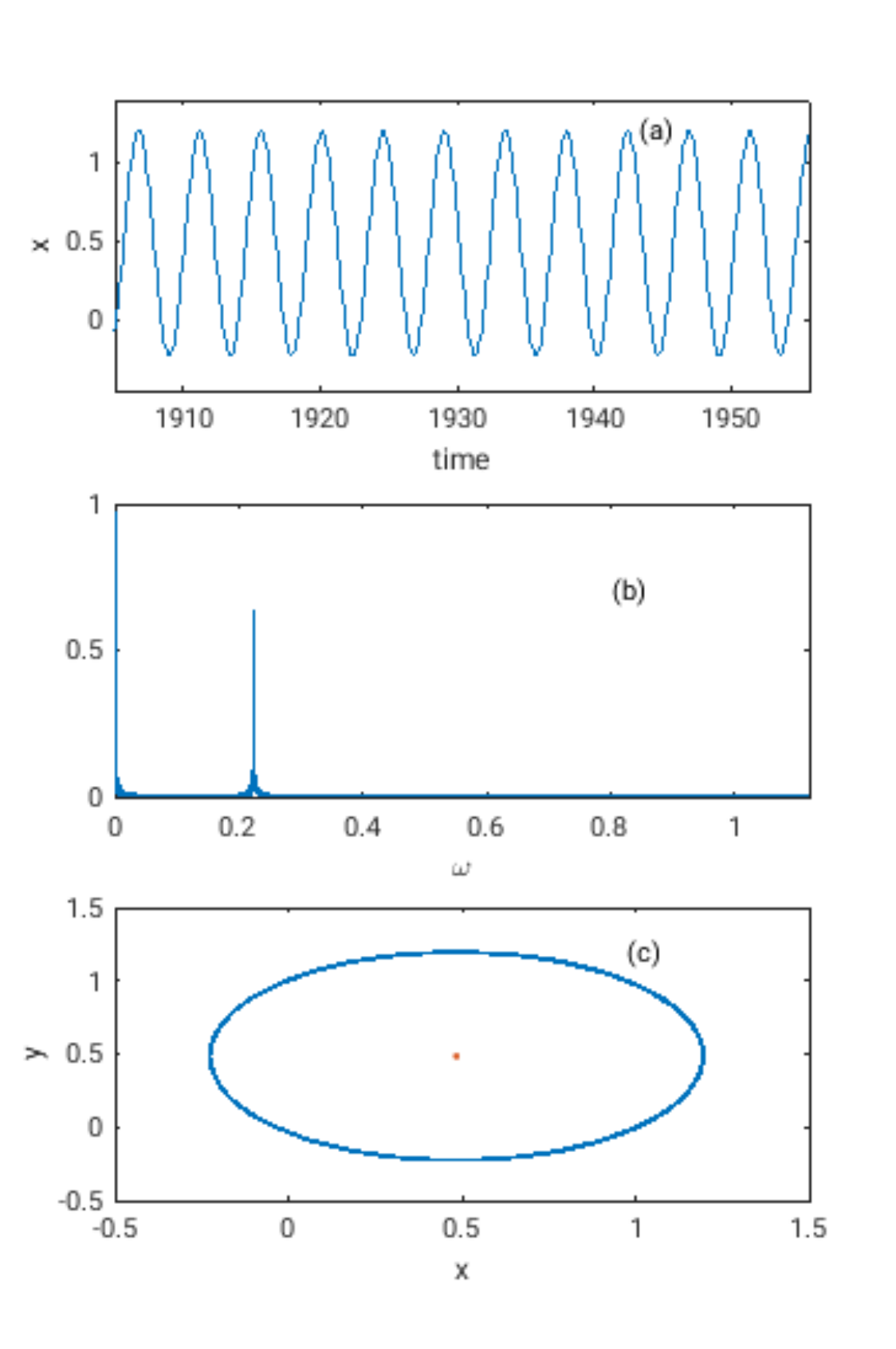}
    \caption{Swarm ring state for $\epsilon = 0.01, a=2.0, \tau = 1.75$. (a)Time
      series of the x-component of a single agent. (b)The power spectrum
      showing a sharp frequency. (c)A phase portrait of the orbit of a single
      agent. The red point denotes the center of mass.}
    \label{Figeps0p01}
  \end{figure}
\end{center}

Note that even when $\epsilon$ is very small, as shown in Fig.~\ref{ThreeSnaps}, we
observe a mix of clustered states which are a combination of  pure ring and
rotation states. The agents tend to cluster into local groups, and the
  clusters move in clockwise and counter-clockwise directions as in the ring state.  Here, however, the phase differences between agents are
non-uniform. When examining a
single random agent, as shown in Fig~\ref{Figeps0p01},  it is periodic with a sharp frequency of
rotation, and the relative positions of all individual agents are phase locked.   When considering the
center of mass  of the positions over all agents, $\mb{R} \equiv \frac{1}{N}
\sum_i \mb{r}_i$, the center of mass 
does small amplitude oscillations about a fixed point (not shown).

\begin{center}
  \begin{figure}[t]
    \includegraphics[width=0.35\textwidth]{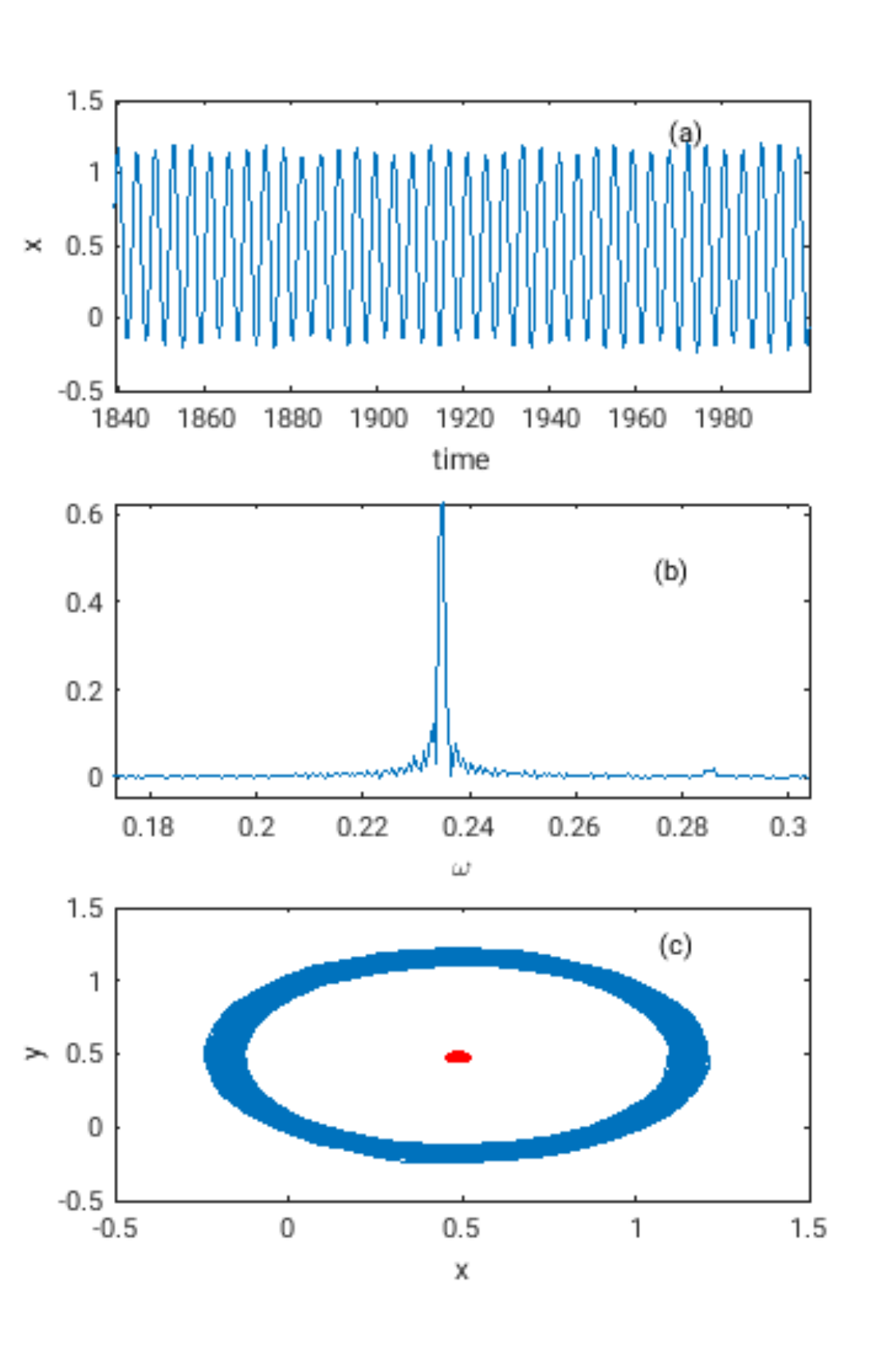}
    \caption{Swam instability $\epsilon = 0.25, a=2.0, \tau = 1.75$. (a)Time
      series of the x-component of a single agent. (b)The Power spectrum
      showing a slight broadening and birth of a new frequency. (c)A phase
      portrait of the orbit of a single agent.}
    \label{Figeps0p25}
  \end{figure}
\end{center}

\begin{center}
  \begin{figure}[b]
    \includegraphics[width=0.35\textwidth]{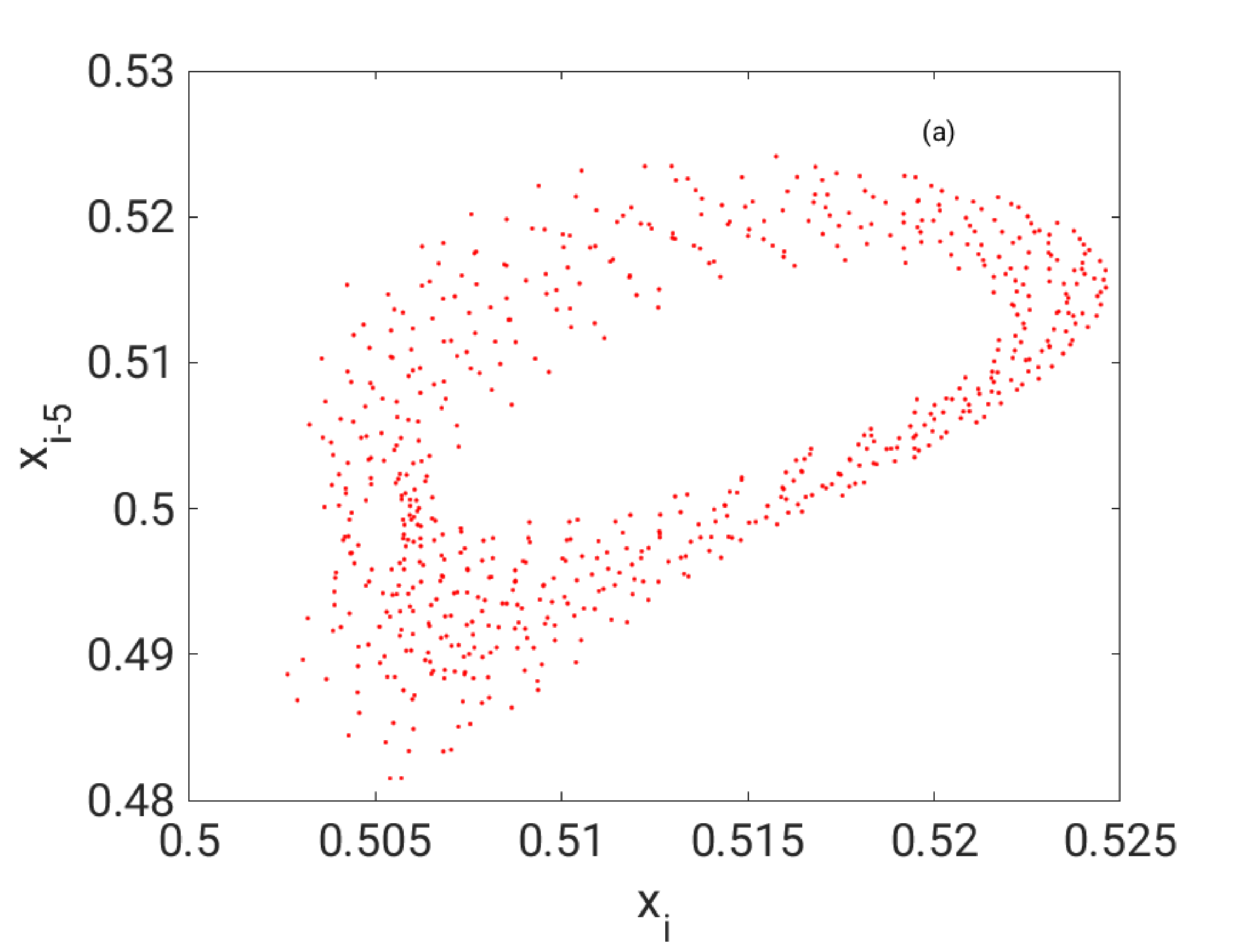}\\ \includegraphics[width=0.35\textwidth]{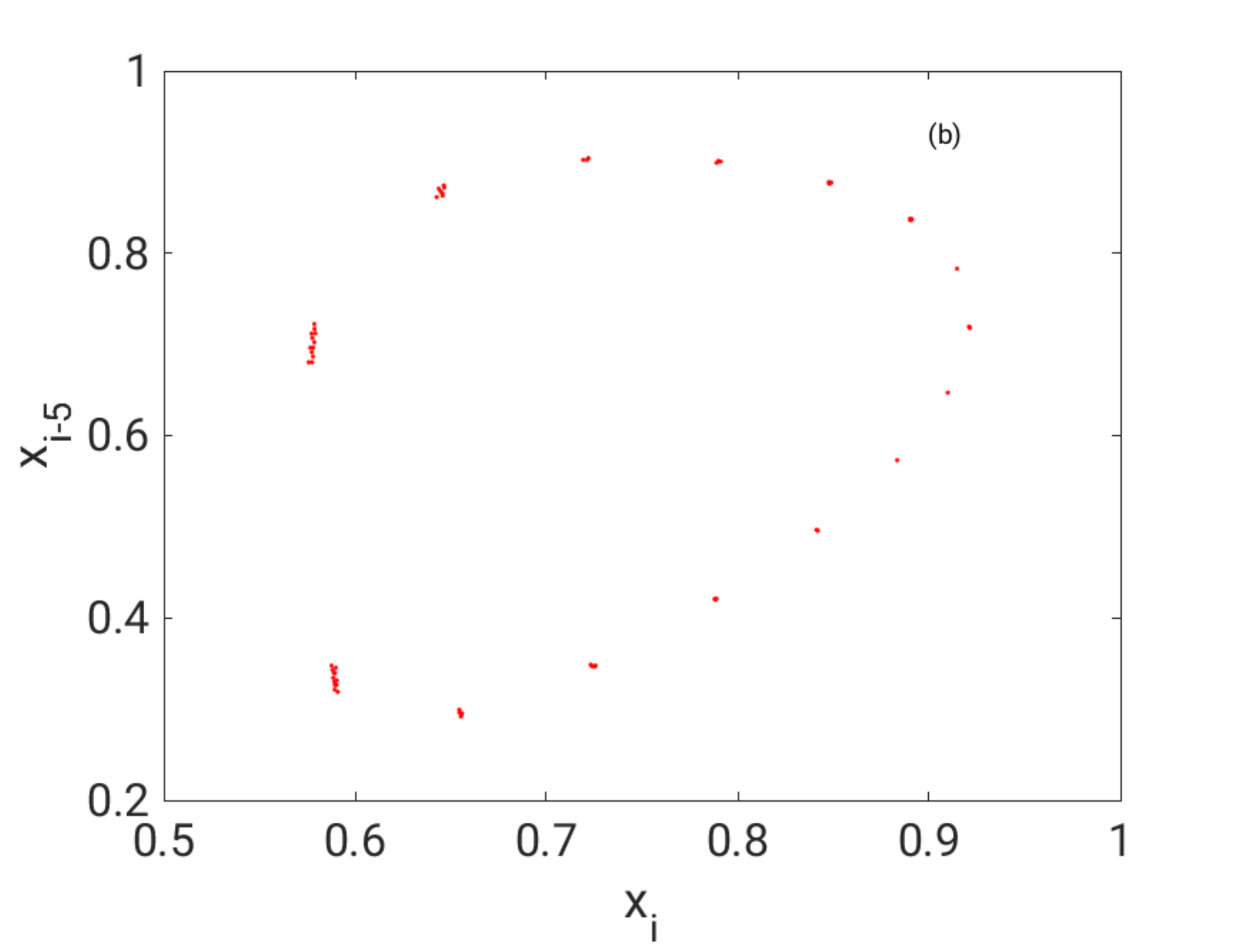}
    \caption{Poincare map of Eqs.~\ref{Eq:agenti}-\ref{Eq:Ci} for (a) $\epsilon = 0.25$, (b) $\epsilon=0.5$. Other
      parameters are fixed: $ a=2.0, \tau = 1.75$. See text for details.}
    \label{Fig:PM}
  \end{figure}
\end{center}

As the radius $\epsilon$ increases, instability of the periodic mixed  state 
occurs, giving rise to more complicated behavior, as seen in
Fig.~\ref{Figeps0p25}. New frequencies are introduced, causing the ring state to
appear as a quasi-periodic attractor. Moreover, the  dynamics of the
center of mass  has its own non-trivial
dynamics which includes the effects of new frequencies.  {By examining
the Poincare map of the attractors, the instability gives rise to dynamics which we conjecture is
motion on a 
torus.} Letting $(M_x, M_y)$ denote the time
averaged center of mass over all agents, we compute the sequence ${x(t_i),
  i=1..M}$ when $y(t_i)=0$ and $x(t_i)>M_x$. The  result is shown in the two
panels in Fig.~\ref{Fig:PM}. Panel (a) shows a complicated toroidal motion
after transients are removed of the center of mass in
  Fig.~\ref{Figeps0p25}c. For a single frequency, the dynamics of the center of
  mass would be a single fixed point. The addition of new frequencies is
  revealed in the Poincare map as complicated motion on a torus. For larger values of $\epsilon$, the motion on
the torus converges to a periodic attractor in panel (b).

\section{Mean-Field Equation of Range Dependent Delay Coupled Swarm} 
In order to shed some light on the origin of the bifurcation to dynamics on a
torus, we examine the full swarm model from a mean-field perspective. The mean
field is much lower dimensional, and a full bifurcation analysis may be
done. We consider the case of all-to-all communication.  Let

\[
\bR = \frac{1}{N}\sum_{i = 1}^N\mathbf{r}_i 
\]
and 
\[
\mathbf{r}_i  = \bR + \delta \mathbf{r}_i, 
\]
where $\delta\mathbf{r}_i$ is a fluctuation term with the identity, and  
\begin{equation}
\sum_{i = 1}^N \delta \mathbf{r}_i  = 0. \label{fruc_def}
\end{equation}
Then we can write Eq.~\ref{Eq:agenti_range}  as 
\begin{align} \label{Eq:MFsub}
\ddot{\bR} + \delta\ddot{\br}_i & = ( 1 - |\dot{\bR} + \delta \dot{\br}_i|^2)(\dot{\bR} + \delta \dot{\br}_i)\nonumber\\
& - \frac{a}{N}\sum_{j = 1, j\neq i}^N((\bR + \delta\mathbf{r}_i) - (\bR^{\tau}
                                                                                                                 + \delta\mathbf{r}_j^{\tau}))C_{1,i}\nonumber \\& - \frac{a}{N}\sum_{j = 1, j\neq i}^N((\bR + \delta\mathbf{r}_i) - (\bR + \delta\mathbf{r}_j))C_{2,i}, 
\end{align}
where
\begin{align*}
C_{1,i} & = H(\lVert{\mathbf{r}_i - \mathbf{r}_j^{\tau}\rVert} - \epsilon)\\
& = H(\lVert{(\bR + \delta\mathbf{r}_i )- (\bR^{\tau} + \delta\mathbf{r}_j^{\tau})\rVert} - \epsilon)\\
& = H(\lVert{\bR -\bR^{\tau} + \delta\mathbf{r}_i  - \delta\mathbf{r}_j^{\tau}\rVert} - \epsilon)
\end{align*}
and 
\begin{equation*}
C_{2,i} =1 -  C_{1,i} .
\end{equation*}
We use the following  to reduce the equations of motion to the mean
field: From Eq.~\ref{fruc_def}, we note  
\begin{align}
\sum_{i = 1}^N \delta \mathbf{r}_i^{\tau} & = \sum_{j = 1, j\neq i}^N\delta\mathbf{r}_j^{\tau} + \delta\mathbf{r}_i^{\tau} = 0 \iff\nonumber\\
&-\sum_{j = 1, j\neq i}^N\delta\mathbf{r}_j^{\tau}  = \delta\mathbf{r}_i^{\tau} \label{trick}.
\end{align}

We further assume that all perturbations from the mean,  $\delta\mathbf{r_i}$,
are all negligible. (This is always true if the coupling amplitude is
sufficiently large.) In addition, we use the fact that  $\DS{\frac{a(N-1)}{N}}$ limits to $a$, as $N\to\infty$. Therefore, we obtain mean field approximation for the center of mass of range dependent coupled delay case:
\begin{equation}
\ddot{\bR} = ( 1 - \lvert\dot{\bR}\rvert^2)\cdot\dot{\bR} \label{MF_eqn}
- a(\bR - \bR^{\tau})\cdot H(\lVert{\bR -\bR^{\tau} \rVert} - \epsilon)
\end{equation}  

\section{Numerical Analysis of the mean field equation}

\subsection{Examples of rotational attractors}
\begin{center}
  \begin{figure}
    \includegraphics[width=0.35\textwidth]{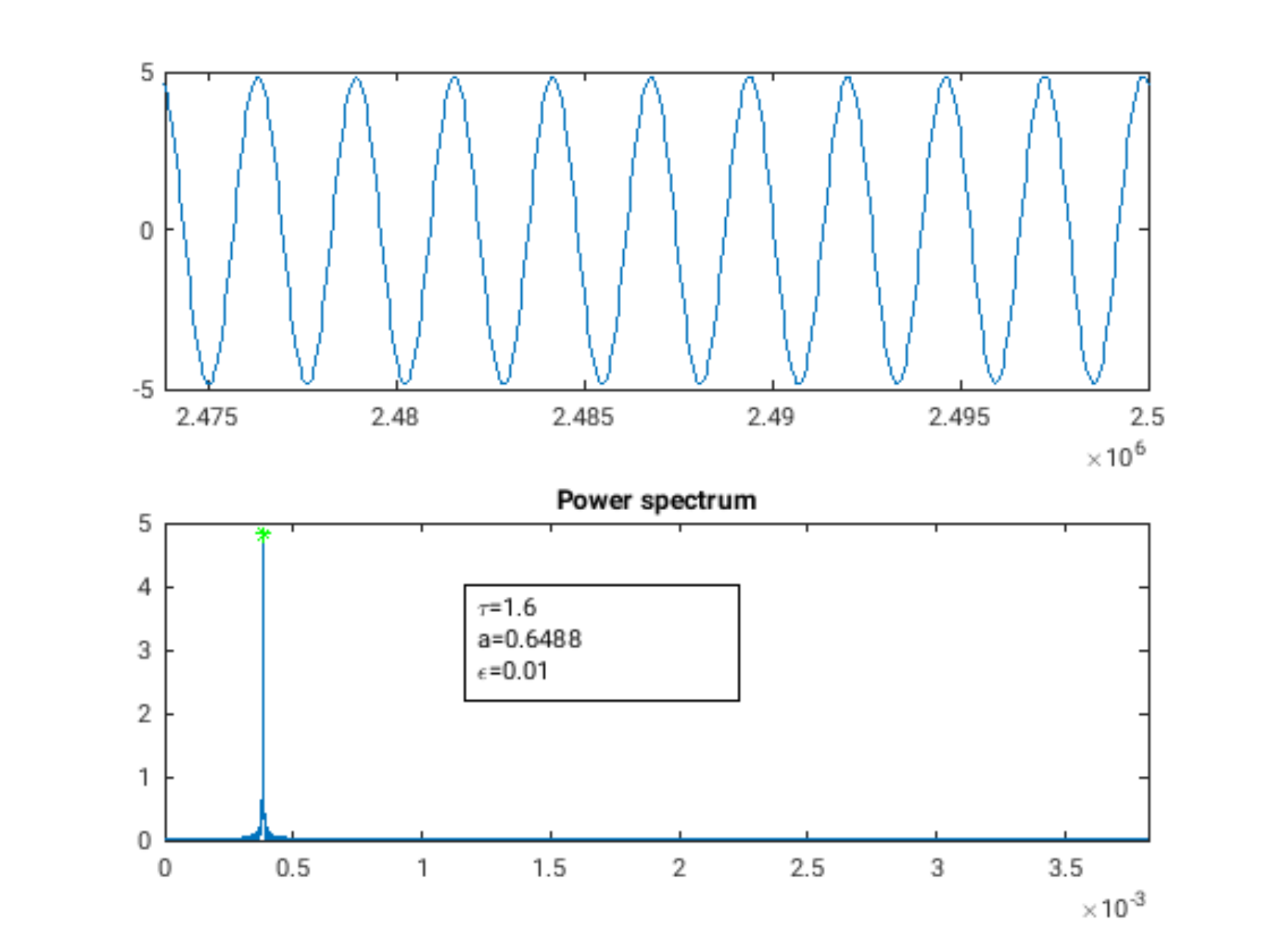}
    \caption{Periodic motion of the mean field Eq.~\ref{MF_eqn} for $\epsilon=0.01, a=0.64, \tau=1.6$. (a) Time
      series of the x-component of the mean field. (b) Power spectra of the
      time series. }
    \label{Fig:PS_small_eps}
  \end{figure}
\end{center}

\begin{center}
  \begin{figure}[t]
    \includegraphics[width=0.35\textwidth]{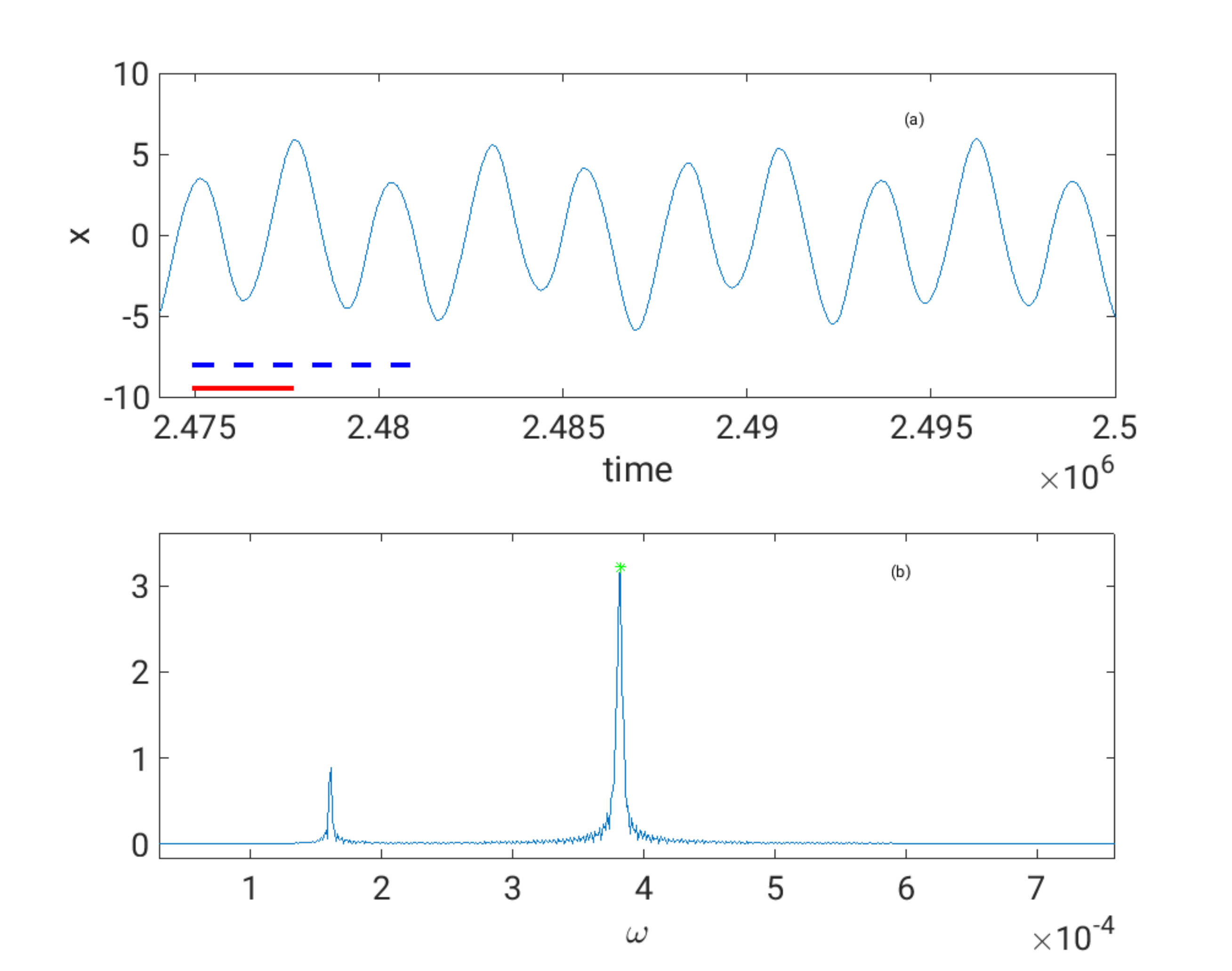}
    \caption{Quasi-periodic motion of the mean field Eq.~\ref{MF_eqn}.  (a) Time
      series of the x-component of the mean field. Solid (red) line denotes
      period length of dominant spectral peak. Dashed line denotes period
      length of secondary peak.  (b) Power spectra of the
      time series. }
    \label{Fig:PS_large_eps}
  \end{figure}
\end{center}

As in the case for the full multi-agent system, we see
the existence of periodic behavior for $\tau$ sufficiently below an
instability threshold, as shown in
the time series of Fig.~\ref{Fig:PS_small_eps}. As we increase $\tau$, we expect
the periodic orbit to lose stability, resulting in a new attractor. In
particular, one notices the emergence of a new frequency in addition to the
existing dominant one, as shown in Fig.~\ref{Fig:PS_large_eps} The additional frequency usually implies a bifurcation to
dynamics on a torus, or a higher dimensional torus. 

We now investigate this transition by tracking the stability via monitoring the
Floquet exponents corresponding to the periodic orbit. For a general
differential delay equation given by $\dot{\bx}(t)=\bF (\bx (t),\bx
(t-\tau))$, if $\bm{\phi}(t) = \bm{\phi}(t+T)$ for all $t \ge 0$, then
stability is determined by examining the linearized equation along
$\bm{\phi}(t)$:
\begin{align}\label{Eq:LVE}
\dot{\bm{X}}(t) &= \frac{\partial \bF}{\partial
  \bx(t)}(\bm{\phi}(t),\bm{\phi}(t-\tau))\bm{X}(t)\nonumber  \\ &+ \frac{\partial \bF}{\partial \bx(t-\tau)}(\bm{\phi}(t),\bm{\phi}(t-\tau))\bm{X}(t-\tau).
\end{align}
The stability of the periodic solution is determined by the spectrum of the
time integration operator $U(T, 0)$ which integrates Eq.~\ref{Eq:LVE}
around $\phi (t)$ from time t = 0 to t = T. This operator is called the monodromy
operator and its (infinite number of) eigenvalues, which are independent of the
initial state, are called the Floquet multipliers \cite{Hale1977}.
For autonomous systems, it is necessary and sufficient there exists a trivial Floquet multiplier at 1,
corresponding to a perturbation along the periodic solution \cite{Hartungetal2006,Hale1993}. The periodic solution is stable provided all multipliers (except the trivial one) have modulus
smaller than 1; it is unstable if there exists a multiplier with modulus larger
than 1. Bifurcations occur whenever Floquet multipliers move into or out of
the unit circle. Generically three types of bifurcations occur in a one parameter
continuation of periodic solutions: a turning point, a period doubling,  and a torus bifurcation where a branch
of quasi-periodic solutions originates and where a complex pair of multipliers
crosses the unit circle \cite{Hale1977}.

\begin{center}
  \begin{figure}[t]
    \includegraphics[width=0.35\textwidth]{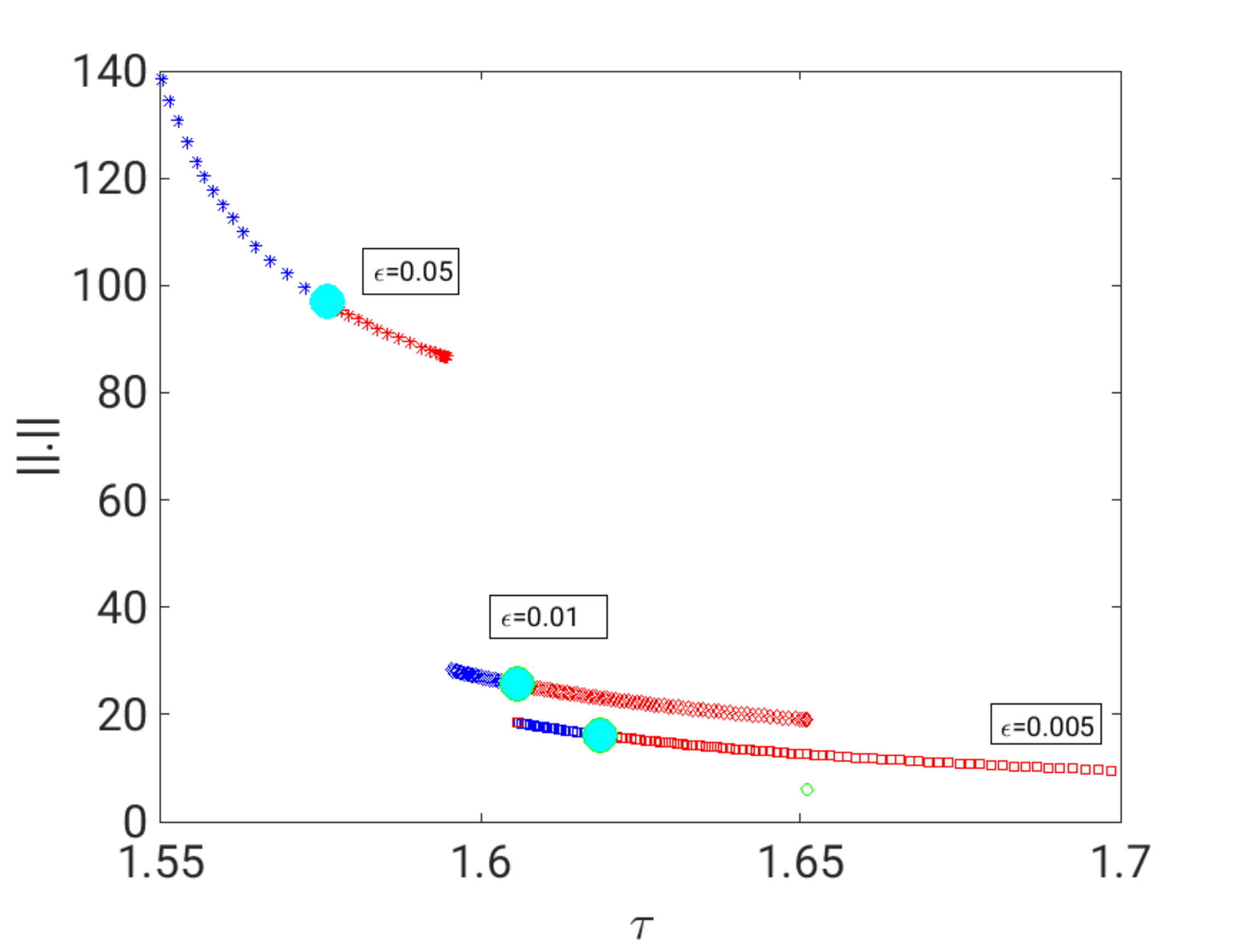}
    \caption{Bifurcation plot showing the norm of the periodic orbits as a
      function of delay $\tau$. Parameter a=0.68. Red (blue) markers denote
      unstable (stable) orbits.Cyan symbols denote the change in stability
      where a pair of complex eigenvalues cross the imaginary axis.   }
    \label{Fig:Bif_tau}
  \end{figure}
\end{center}

We have tracked a set of stable periodic orbits for various radii of
$\epsilon$, and located the change in stability by computing the Floquet multipliers. The results plotted in
Fig.~\ref{Fig:Bif_tau} show that for a range of radii $\epsilon$, there exists a
bifurcation to a torus at some delay. Notice that as $\epsilon$ increases, there results an
increase in the size of the orbits, which qualitatively agrees with our full
agent based simulations.

\begin{center}
  \begin{figure}
    \includegraphics[width=0.35\textwidth]{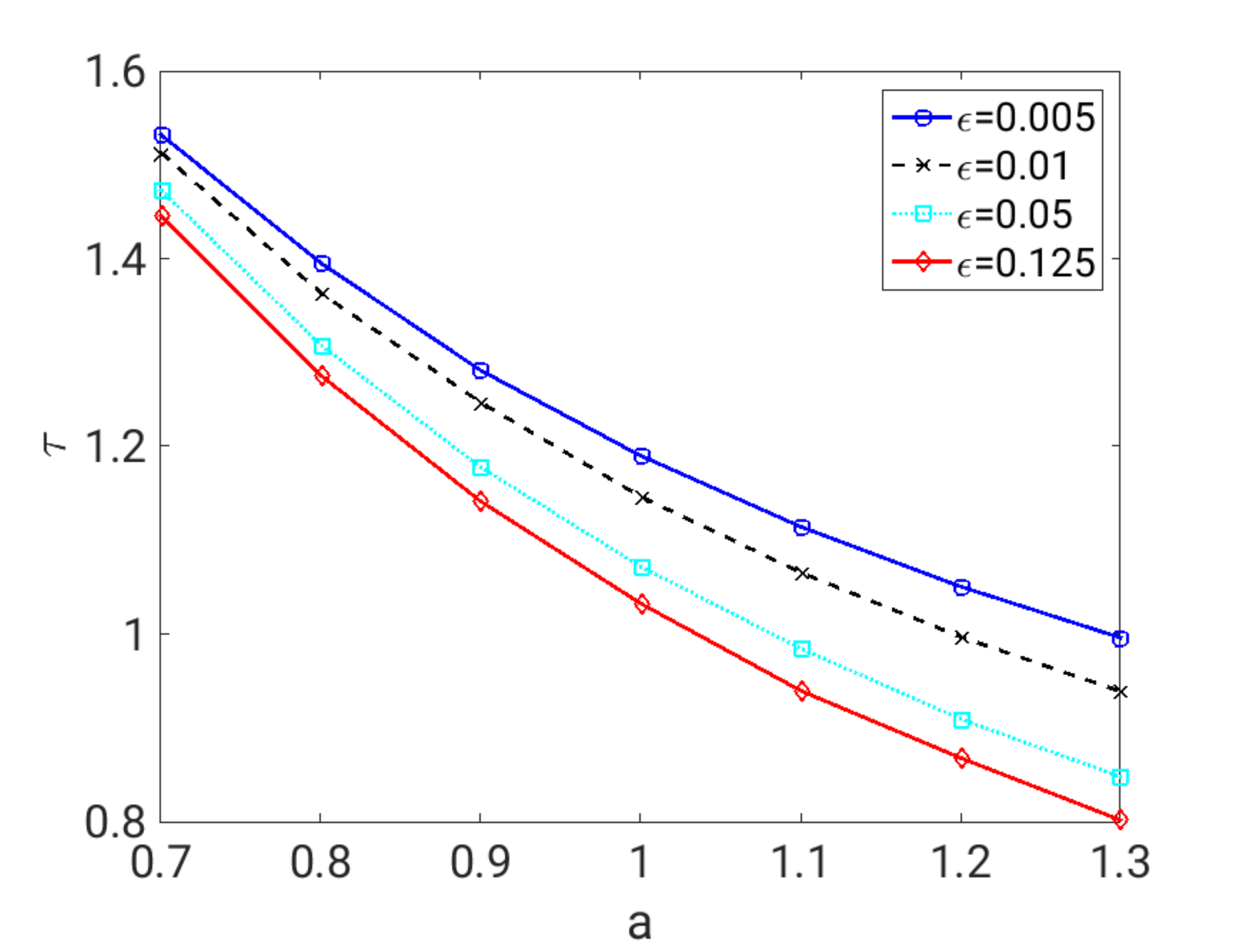}
    \caption{Plotted is the locus of points at which torus bifurcations emerge
      as a function of coupling amplitude $a$, delay $\tau$ for various range
  radii $\epsilon$ for the mean field Eq.~\ref{MF_eqn}.}
    \label{Fig:TB_comp}
  \end{figure}
\end{center}

Since there exists a range of delays which destabilize periodic swarm dynamics
for each $\epsilon$, we summarize the onset of torus bifurcations by plotting
the locus of points at which stability changes as a a function of coupling
amplitude and delay. The results are plotted in Fig.~\ref{Fig:TB_comp}.

Figure~\ref{Fig:TB_comp} is revealing in that it shows a functional
relationship of the bifurcation onset that is similar over a range of
$\epsilon$. For larger values of $\epsilon$, it is clear that lower values of
delay and coupling are required to generate bifurcations. This holds true
over two orders of $\epsilon$. For a fixed value of $\epsilon$, we also see
monotonic relationship between delay and coupling strength, so that it is
easier for smaller delays to destabilize periodic motion for larger coupling
strengths. 

\section{Conclusions}
We considered a new model of a swarm with  delay coupled communication
network, where the delay is considered to be range dependent. That is, given a
range radius, delay is on if two agents are outside the radius, and zero
otherwise. The implication is that small delays do not matter if the agents
are close to each other.

The additional range dependence creates a new set of
bifurcations not previously seen. For general swarms without delay, the
usual states consist of flocking (translation) or ring / rotational state
(milling), with agents spread in phase. With the addition of a fixed delay, a
rotational state bifurcates that has all agents in phase and rotate
together~\cite{hindes2020unstable}. Range dependence introduces a new rotational
bifurcating state that exhibits behavior observed as a new mixed state
combining dynamics of both ring and rotating states.

The radius parameter $\epsilon$, was used to quantify the bifurcation of
the rotational mixed state. For small $\epsilon$, we see dynamics for the full
swarm shows clustered counter-rotational behavior that is periodic. This
agrees for small radius values in the mean field description as well. As the
radius increases, the mixed periodic state generates new frequencies in the
full model, which are manifested as torus bifurcations in the mean
field. Mean field analysis was done by tracking Floquet multipliers that
cross the imaginary axis as complex pairs. Frequency analysis explicitly shows
the additional frequencies in the mean field.

Finally, we tracked the locus of coupling amplitudes and delay for various
values of $\epsilon$ locating the parameters at which torus bifurcation
occur. The results reveal that as $\epsilon$ increases, torus bifurcations
onset at lower values of coupling amplitude and delay.  The implications are
that more complicated behavior than periodic motion has a greater probability of
being observed in both theory and experiment if range dependence of delay is included.

\section{Supplementary Material}

The videos show the attractor of a swarm consisting of N=300 agents.
Fixed parameters for the three videos are $a=2.0, \tau = 1.75$
The
parameters for zero radius (delay is on all the time)  are $\epsilon =
0.0,c_r=0.05$, and $l_r=0.05$ for a baseline, are shown in Video1\_eps\_0p0.mp4.

The parameters corresponding to Fig. 2 are $\epsilon = 0.01,c_r=0.01$, and
$l_r=0.05$ are shown in Video2\_eps\_0p01.mp4.
The video shows that the attractor persists when repulsive forces are local
and weak. Similar behavior is observed when N=150, which is used in
Fig.~\ref{ThreeSnaps} without repulsion; i.e., $c_r=0$.
The parameters for corresponding to Fig. 3 are $\epsilon = 0.25,c_r=0.05$, and
$l_r=0.05$, shown in Video3\_eps\_0p25.mp4.

\begin{acknowledgements}
IBS, JH, IT and KK  gratefully acknowledge ONR for their
support under N0001412WX20083, N0001420WX00034, and the NRL Base
Research Program N0001420WX00410. VE is supported under the  NRL Karles
Fellowship Program, JON 55-N2Q4-09. SK was supported through the GMU Provost
PhD award as part of the Industrial 
Immersion Program. MAH is supported by  ONR  No. N00014-18-1-2580 and ARL DCIST CRA W911NF-17-2-0181.

\end{acknowledgements}
  
Data sharing is not applicable to this article as no new data were created in this study.


%

\end{document}